# Ferromagnetic GaMnAs/GaAs superlattices – MBE growth and magnetic properties

*J. Sadowski[a, b, c], R. Mathieu[d], P. Svedlindh[d], M. Karlsteen[a], J. Kanski[a], Y. Fu[a]

J. T. Domagała[b], W. Szuszkiewicz[b], B. Hennion[e], D. K. Maude[f], R. Airey[g], G. Hill[g]

*(a) Department of Experimental Physics, Chalmers University of Technology*

*and Göteborg University, SE-412 96 Göteborg, Sweden*

*(b) Institute of Physics Polish Academy of Sciences, al. Lotnikòw 32/46,*

*PL-02-668 Warszawa, Poland*

*(c) Oersted Laboratory, Copenhagen University, Denmark*

*(d) Uppsala University, Department of Materials Science, SE-751 21 Uppsala, Sweden*

*(e) Laboratoire Leon Brillouin, CE Saclay, 91191 Gif-sur-Yvette, France*

*(f) Grenoble High Magnetic Field Laboratory, 38042 Grenoble, France*

*(g) ) Department of Electronic and Electrical Engineering, University of Sheffield, U.K*

---

*

We have studied the magnetic properties of $(GaMnAs)_m(GaAs)_n$ superlattices with magnetic GaMnAs layers of thickness between 8 and 16 molecular layers (ML) (23–45 Å), and with nonmagnetic GaAs spacers from 4 ML to 10 ML (11–28 Å). While previous reports state that GaMnAs layers thinner than 50Å are paramagnetic in the whole Mn composition range achievable using MBE growth (up to 8% Mn), we have found that short period superlattices exhibit a paramagnetic-to-ferromagnetic phase transition with a transition temperature which depends on both the thickness of the magnetic GaMnAs layer and the nonmagnetic GaAs specer. The neutron scattering experiments have shown that the magnetic layers in superlattices are ferromagnetically coupled for both thin (below 50 Å) and thick (above 50 Å) GaMnAs layers.

1. **Introduction**

III-V ferromagnetic semiconductors such as InMnAs [1–5] and GaMnAs [6–11] have been gaining an increasing interest in recent years due to the expanding research activity in the field of spintronics [12], where the possibility of integrating spin dependent effects into the electronic devices seems to be very attractive. In this context, low dimensional structures like superlattices, quantum wires and dots incorporating ferromagnetic semiconductor materials are particularly interesting. So far, mainly 2D structures based on III-V ferromagnetic semiconductors have been investigated in detail, however the understanding of their properties is far from being complete. In previous studies of superlattices (SL) with magnetic GaMnAs layers and nonmagnetic AlGaAs and InGaAs spacers the ferromagnetic properties were reported to disappear when the thickness of the individual magnetic layers decreased below 50 Å [13-18]. On the one hand, recent theoretical modelling of magnetism in GaMnAs quantum well structures [19], based on Monte-Carlo simulations on a confinement-adapted RKKY exchange interaction model, predicts a lower limit for the GaMnAs thickness of 9 ML (25 Å) for the ferromagnetic phase to occur. On the other hand, theoretical modelling of magnetism in GaMnAs/GaAs SL structures found no evidence of a corresponding GaMnAs thickness limit [20]. Moreover, Jungwirth et al. [20] predict an oscillating character of the magnetic coupling between 7ML thick GaMnAs layers in GaMnAs/GaAs SL structures - antiferromagnetic or ferromagnetic depending on the GaAs spacer layer thickness.

Our results show that GaMnAs/GaAs superlattices have a paramagnetic-to-ferromagnetic phase transition for a magnetic layer thickness of 8 molecular layers, i.e. 22.8 Å, which is much below the 50 Å thickness limit previously reported [13 – 18]. We have also confirmed using neutron scattering experiments, that for a GaAs spacer thickness of 4 to 6 ML there is a ferromagnetic coupling between GaMnAs layers [21].

## 2. MBE growth and properties of single GaMnAs layers

GaMnAs samples were grown in a MBE system (KRYOVAK) equipped with an $As_2$ valve cracker source. As substrates we used Semi-insulating, epi-ready GaAs(100) wafers. The MBE growth of superlattice structures was preceded by a standard high-temperature (HT) growth of a 5000 Å thick GaAs buffer. Subsequently the substrate temperature was decreased to 200 °C and stabilized for 1 hour. Afterwards, SL structures were grown with 30 s growth interruptions at each interface. The number of repetitions of the GaMnAs/GaAs sequence was between 100 and 250 depending on the type of the structure.

The substrate temperature during MBE growth was close to 200 °C, and the $As_2$ versus Ga flux ratio was chosen to be about 2. Before the MBE growth of each GaMnAs/GaAs SL, the growth rates of GaAs and GaMnAs were carefully calibrated using reflection high energy electron diffraction (RHEED) intensity oscillations. Analyzing RHEED oscillations recorded during growth of different SL layers we have also calibrated in-situ the Mn composition in GaMnAs. Because the sticking coefficients of Ga and Mn atoms are equal to 1 at low growth temperature, the Mn content in GaMnAs layers is proportional to the growth rate increase of GaMnAs with respect to GaAs. Figure 1 shows RHEED oscillations recorded during growth of a SL structure having 16 monolayers (ML) of $Ga_{0.93}Mn_{0.07}As$ and 9 ML of GaAs.

As reported by several groups [22 – 24], the properties of GaMnAs depend strongly on the LT MBE growth conditions, such as substrate temperature ($T_s$) and As to Ga flux ratio. The differences in $T_s$ as well as in As/Ga give different values of the lattice parameter of GaMnAs [24, 25]. In our case: $T_s$ = 200 °C and $As_2/Ga$ = 2, and the relaxed (unstrained) lattice constant of GaMnAs increases with the Mn content according to the formula [26]:

(1) $a(Ga_{1-x}Mn_xAs) = 5.65469 * (1-x) + 5.9013 * x$

This gives the lattice constant of LT GaAs as $a_{LTGaAs}$ = 5.65469 Å, and that of a hypothetical zinc-blende MnAs as $a_{MnAs}$ = 5.9013 Å. The first value, $a_{LTGaAs}$, is slightly higher than the lattice parameter of bulk GaAs, as well as GaAs epilayers grown at standard, high substrate temperatures (500 °C – 600 °C), which is equal to 5.65333 Å. The lattice parameter expansion of LT GaAs is due to the presence of As antisites [27 - 29], which are incorporated at the concentrations up to 0.5% [29] under LT MBE growth conditions. As it has been previously found [28, 30, 31], the lattice expansion depends linearly on the concentration of As antisites, implying that the lattice parameter of LT GaAs deduced from equation (1), as well as verified by X-ray diffraction (XRD) measurements of LT GaAs layers grown under the same conditions as GaMnAs, gives the concentration of As antisites (As[Ga]) in LT GaAs. The lattice constant $a_{LTGaAs}$ taken from equation (1) corresponds to a concentration of As antisites equal to $3 \times 10^{19}$ cm$^{-3}$.

The $a_{MnAs}$(zinc-blende) = 5.903 Å lattice parameter is lower than the value of 5.98 Å most frequently cited by other authors [6, 8, 9]. It was confirmed recently by Schott et al. [24] that $a_{MnAs}$(zinc-blende) extrapolated from the $a_{Ga(1-x)Mn(x)As}$ vs. x relation depends on the MBE growth temperature ($T_s$), giving 5.90 Å for $T_s$ = 220 °C and 5.98 Å in the case of $T_s$ = 270 °C. All GaMnAs samples studied in this work were grown at low substrate temperature ($T_s$ = 200 °C) and the results of Schott et al. are thus in accordance with our findings.

Theoretical work have shown that As antisites play an important role in GaMnAs [32, 33], strongly influencing the magnetic properties of this material. Sanvito et. al. [32] found that not only the total concentration of antisites, but also their position with respect to the Mn ions are relevant for the magnetic properties of GaMnAs.

For the electronic structure of GaMnAs, the donor-like character of As[Ga] is essential. As is well known [6 – 9], Mn is an acceptor in GaMnAs. However, the hole densities in GaMnAs, measured directly [34] or calculated within the mean-field model using experimental values of the ferromagnetic phase transition temperatures [35], are only a small fraction (from 10% to 30%) of the concentration of Mn ions. The missing holes can be associated with charge carrier compensation by As antisites, which form deep donor states. However, since the antisite density in our GaMnAs films is estimated to be 4 x $10^{19}$ cm$^{-3}$ or smaller, it means that not more than ~ $10^{20}$ cm$^{-3}$ holes can be compensated by these defects, whereas an order of magnitude higher compensation would be necessary to explain the much lower than expected hole concentration. Other mechanisms based on hole localization [35, 36] have been proposed to explain the low value for the hole concentration, but no definite explanation has been accepted so far.

3. Hole density distribution

A high density of holes is a crucial factor for the ferromagnetic phase to occur in GaMnAs as well as in other ferromagnetic semiconductors [10, 36-39], as is described theoretically by several models [10, 37, 40-41], and being well established experimentally [22, 23, 42]. We have measured the density of holes in 500 Å thick Ga$_{0.945}$Mn$_{0.055}$As reference layers grown at the same conditions (As$_2$/Ga flux and substrate temperature) as the GaMnAs/GaAs SL structures. Figure 2 shows the Hall voltage vs. magnetic field, measured at 70 mK and for magnetic field intensities up to 22 T. In the low field region, shown in detail in the right inset, the Hall voltage traces the sample magnetization (anomalous Hall effect), rising steeply for fields between 0T and 0.2T and then saturating. In the high field region, shown in detail in the left inset, the magnetization is saturated and the ordinary Hall coefficient prevails. The Rxy vs. B slope gives the concentration of holes as 2x10$^{20}$ cm$^{-3}$, which is slightly lower than the value of 3.5 x 10$^{20}$ reported by Ohno et al. [34] for GaMnAs containing 5.3% Mn. This may be connected with the lower growth temperature (200 $^{o}$C) used by us. The dependence of the GaMnAs lattice constant on Mn concentration in our case [cf. Eq. (1)] compared to that established by Ohno et al. [6, 9] (giving a$_{MnAs}$(zinc-blende) = 5.98 Å) suggests, in connection to the observation of Schott et al. [24], that the GaMnAs samples studied by Ohno et al. were grown at a higher temperature than our samples.

Having established the concentration of holes in single GaMnAs layers and knowing the concentration of As antisites in the GaAs spacer layers it is possible to calculate the hole concentration profiles in the GaMnAs/GaAs SL structures by solving the one-dimensional Schrödinger equation self-consistently with the Poisson equation [43]. The results of such calculations for SL structures with 8ML GaMnAs and two different GaAs spacer thickness: 4 ML and 10 ML, are shown in Fig. 3. A 50% difference in the maximum concentration of holes in these two structures can be seen. The average hole concentration integrated over the 8 ML thick GaMnAs layer in the structure with 10 ML thick GaAs spacers decreases significantly in comparison to situation in the SL structure with 4 ML thick spacers. It should be noted however, that our calculations are valid only above Tc (i.e. in the paramagnetic state). Below Tc, when the Mn ions are ferromagnetically coupled, a strongly asymmetric distributions of spin-up and spin-down holes are predicted [44]. This asymmetry should lead to an enhanced Tc since the spin down holes mediating the ferromagnetic interactions between Mn ions are confined in the GaMnAs layers. The simple model calculations used here are thus not expected to give quantitative information on the hole concentration, but are used to give the qualitative behaviour of the charge carrier concentration profiles in GaMnAs/GaAs SLs and how it develops as, for instance, the spacer layer thickness is varied.

For the set of samples with 8 ML, 12 ML and 16 ML thick GaMnAs, we observe that the ferromagnetic phase transition disappears for structures with GaAs spacers thicker than 10 ML. One possible reason for this may be the above mentioned decrease of hole concentration in GaMnAs with increasing GaAs spacer thickness. However, as can be seen in Fig. 3, the average hole density in the GaMnAs layers in the SL with10 ML thick

GaAs spacers is ~ 7 x $10^{19}$ cm$^{-3}$, which according to the mean field model of Dietl et al. [36] should yield a paramagnetic-to-ferromagnetic transition at ~ 10 K in GaMnAs with 7% of Mn. The disappearance of the ferromagnetic phase in structures with GaAs spacers thicker than 10 ML, should thus be caused by some other effect than the low value of the hole density in GaMnAs. One possible origin may be interface imperfections. The interface roughness of about 2 monolayers, may give an effective thickness of the magnetic GaMnAs layers smaller than 8 monolayers. This together with the low concentration of holes may explain the disappearance of the paramagnetic-to-ferromagnetic transition in SLs having a spacer layer thickness being larger than 10 ML.

## 4. Structural and magnetic properties of GaMnAs/GaAs superlattices.

Before the MBE growth of each GaMnAs/GaAs SL, the growth rates of constituent SL layers were carefully calibrated on a test sample via RHEED intensity oscillations. The MBE growth of the SLs was controlled by RHEED oscillations too. The sequence of shutter opening times was chosen in such a way, that the growth of each layer was terminated at the maximum intensity of the specular RHEED beam, as can be seen in Fig. 1. The thickness of the GaMnAs and GaAs layers was thus controlled with a submonolayer accuracy. This is in our opinion a key factor to obtain short period GaMnAs/GaAs SLs structures with ferromagnetic properties. At each interface the growth was stopped for 30s in order to enhance the surface smoothness. Using this growth procedure it was usually possible to record RHEED intensity oscillations for up to 50 superlattice periods.

The structural quality of the SLs was estimated by X-ray diffraction (XRD) measurements. Figure 4 shows an example of XRD results for the SL structure with 12 ML thick GaMnAs layers and 6 ML thick GaAs spacers. The position of the zero-order SL (004) Bragg peak as well as the positions of the satellites of order +1 and −1 are well reproduced by XRD simulations. The parameters extracted for each SL structure, that is the thickness and the composition of each layer in a SL, are in good accordance with parameters measured by RHEED oscillations during the MBE growth. Figure 5 shows the measured XRD spectra for three GaMnAs/GaAs SL structures with 8 ML thick $Ga_{0.96}Mn_{0.04}As$ and GaAs spacers having a thickness ranging from 5 ML to 9 ML in 2 ML steps. It can clearly be seen that the angular spacing of the 1-st order satellites decreases with increasing spacer thickness, which is in agreement with the expected behavior for the XRD spectrum in a system with increasing period. It should be noted that the fact that only first order (and in some cases 2-nd order) satellites can be seen in XRD spectra is due to both low chemical contrast between GaAs and GaMnAs with only up to 7% of Mn, and low lattice parameter contrast (only 0.3% lattice mismatch between $Ga_{0.93}Mn_{0.07}As$ and LT GaAs).

The reciprocal space maps measured by us for some SL structures demonstrate in all cases that the SL structures are totally strained to the GaAs(100) substrate. In relation to this, the GaMnAs/GaAs SL structures are in a compressive strain state, which induces in-plane magnetic anisotropy in this system [9].

Magnetic properties of GaMnAs/GaAs SLs were investigated by means of magnetization measurements using a superconducting quantum interference device (SQUID). Figure 6 shows SQUID results for a $Ga_{0.93}Mn_{0.07}As$/GaAs SL structure with a magnetic layer thickness of 8 ML and 4ML thick GaAs spacer layers. The number of repetitions of the GaMnAs/GaAs sequence was in this case equal to 100. The inset shows magnetization vs. temperature curves for 0.8 µm thick reference GaMnAs films with 5.5% and 7% of Mn. The $T_c$ value close to 60 K for the GaMnAs/GaAs SL is close to the value obtained for the single GaMnAs sample containing 5.5% of Mn. Also, the saturation magnetization measured at 4 K is the same for both samples (~ 2 $\mu_B$/Mn atom). It is interesting to find that $T_c$ for the SL structure with $Ga_{0.93}Mn_{0.07}As$ layers is higher than the transition temperature obtained for the 0.8 _m thick reference $Ga_{0.93}Mn_{0.07}As$ sample. The average Mn concentration in the SL structure, calculated as for a digital alloy, equals 4.67%. A single thick layer of GaMnAs containing 4.67% of Mn has a $T_c$ value slightly lower than that of GaMnAs with 5.5% of Mn (known to exhibit the maximum value of $T_c$ [9]). We have verified by XRD measurements that this superlattice is well defined (i.e. that there is a negligible interdiffusion at the interfaces). It thus appears that the paramagnetic-to-ferromagnetic transition temperature can be enhanced when magnetic GaMnAs layers are separated by non-magnetic GaAs spacers in a superlattice structure. Similar type of behavior has been observed in digital GaAs/MnAs structures [45], where fractional MnAs monolayers (with 50% GaAs surface coverage) are separated by GaAs spacers having a thickness of the same order magnitude as the GaAs spacers in GaMnAs/GaAs SL structures investigated in this work.

Recent theoretical work [44, 46] interpret a $T_c$ enhancement in SLs structures containing GaMnAs as a result of confinement of spin-down holes in the magnetic layers and predict in some specific cases $T_c$ values in SLs to be higher than in single GaMnAs layers [46].

The dependence of the magnetic properties on thickness of the GaAs spacer can be seen in Fig. 7 for GaMnAs/GaAs SLs with 12 ML thick $Ga_{0.96}Mn_{0.04}As$ layers. The magnetic transition temperature decreases monotonously with increasing thickness of the GaAs spacer; for a spacer thickness of 10 ML or larger, it was not possible to detect a low temperature ferromagnetic phase. Similar results with a ~ 10 ML lower limit of the GaAs layer thickness have been obtained for GaMnAs/GaAs SL structures with 8ML and 16 ML thick GaMnAs layers.

Structures with 8 ML, 10 ML, 12 ML and 16 ML thick GaMnAs layers, as well as structures having much thicker GaMnAs layers (25 ML and 50 ML) were investigated by polarized neutron reflectivity [47] and neutron diffraction [21]. In both cases, ferromagnetic coupling between magnetic GaMnAs layers was detected for spacer layer thickness of 6 to 8 monolayers. Figure 8 shows the temperature dependence of the intensity of the 1-st order neutron diffraction satellite for a GaMnAs (10ML) /GaAs (6ML) SL structure with 250 repetitions. An increase in intensity of this satellite diffraction peak below $T_c$ ( $T_c$ was determined by SQUID measurements) proves that the adjacent GaMnAs layers are ferromagnetically coupled [21]. So far, we have not detected the theoretically predicted [20] antiferromagnetic interlayer coupling in this system.

## 5. Conclusions

We have observed a ferromagnetic phase transition in GaMnAs/GaAs superlattices with ultrathin GaMnAs layers (from 8 to 16 ML). This is much below the so far established limit for the GaMnAs layer thickness (50 Å) at which GaMnAs/Ga(Al)As superlattices were reported to loose their ferromagnetic properties. We have verified that the paramagnetic-to-ferromagnetic phase transition occurs in short period GaMnAs/GaAs superlattices with 8 ML, 12 ML and 16 ML GaMnAs layer thickness. In most cases, the ferromagnetic phase in these superlattice disappears when the thickness of the GaAs spacer layer is larger than 10 ML (28 Å). We interpret this 10 ML spacer thickness limit for the ferromagnetic phase in short period GaMnAs/GaAs superlattices as a result of the decrease of hole density in the magnetic layers due to diffusion and partial recombination of holes in the GaAs spacer layers. At low spacer layer thickness (4 ML) the opposite effect occurs – the temperature of the paramagnetic-to-ferromagnetic phase transition in SL structures is higher than in a thick reference GaMnAs sample having the same Mn concentration. The enhancement of $T_c$ in this case can be explained as an effect of confinement of spin-down holes in the GaMnAs layers. Neutron diffraction measurements on structures with 6 ML and 8 ML thick GaAs spacers demonstrate that adjacent GaMnAs layers in the SL structures are ferromagnetically coupled.

## Acknowledgements

This work was supported by grants from the Swedish Natural Science Research Council (NFR), the Swedish Research Council for Engineering Sciences (TFR), and, via cooperation with the Nanometer Structure Consortium in Lund, the Swedish Foundation for Strategic Research (SSF). Measurements at high magnetic fields have been supported by the European Community within the "Access to Research Infrastructure action of the Improving Human Potential Programme".

**Figures**

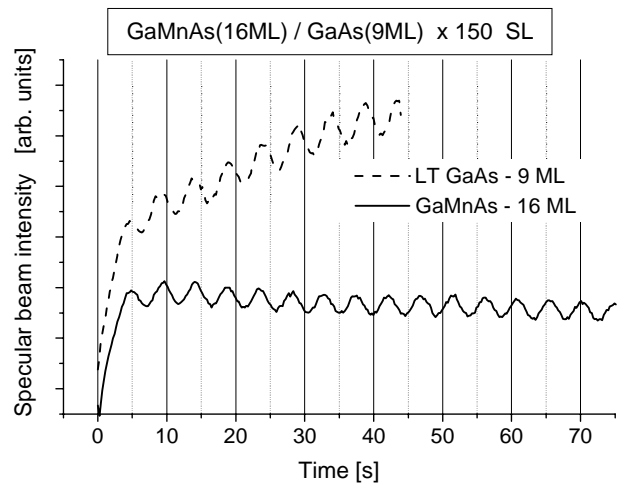

Fig. 1. Reflection High Energy Electron Diffraction (RHEED) intensity oscillations recorded for the specular RHEED beam during the MBE growth of a Ga$_{0.93}$Mn$_{0.07}$As (16 ML) /GaAs (9ML) superlattice structure.

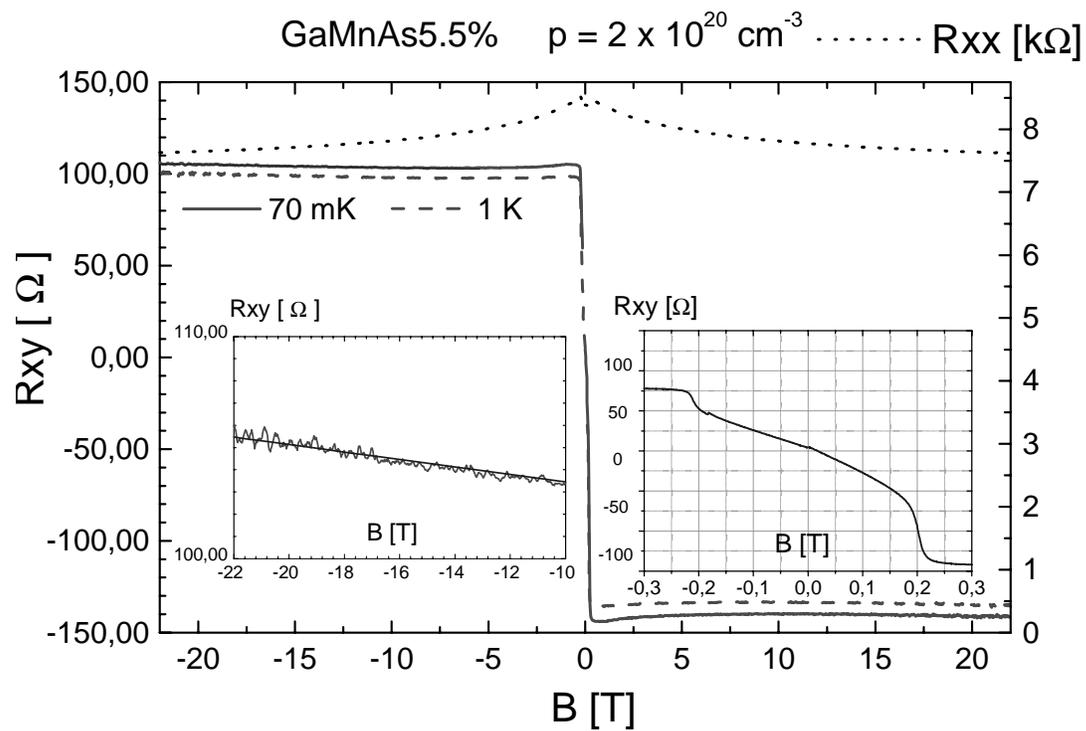

**Fig. 2.** Hall effect results for a 500Å thick $Ga_{0.945}Mn_{0.55}As$ layer capped with 100 Å LT GaAs. The right side inset shows the low field region dominated by the anomalous Hall effect. The left side inset shows the high field region, at which the hole density has been determined from the R(xy) vs. B slope.

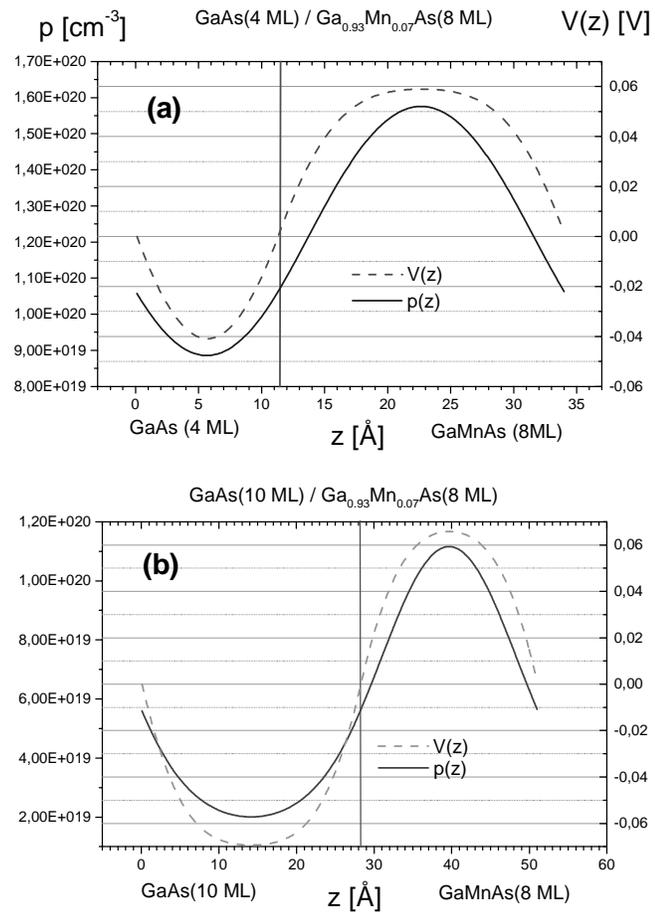

**Fig. 3** Calculated hole densities - p(z) and electrostatical potentails V(z)

in $Ga_{0.93}Mn_{0.07}As$ (m ML) / GaAs (nML) superlattice structures for thickness

of GaMnAs (m) and GaAs (n) being equal to:

(a) m = 8, n = 4,

(b) m = 8, n = 10.

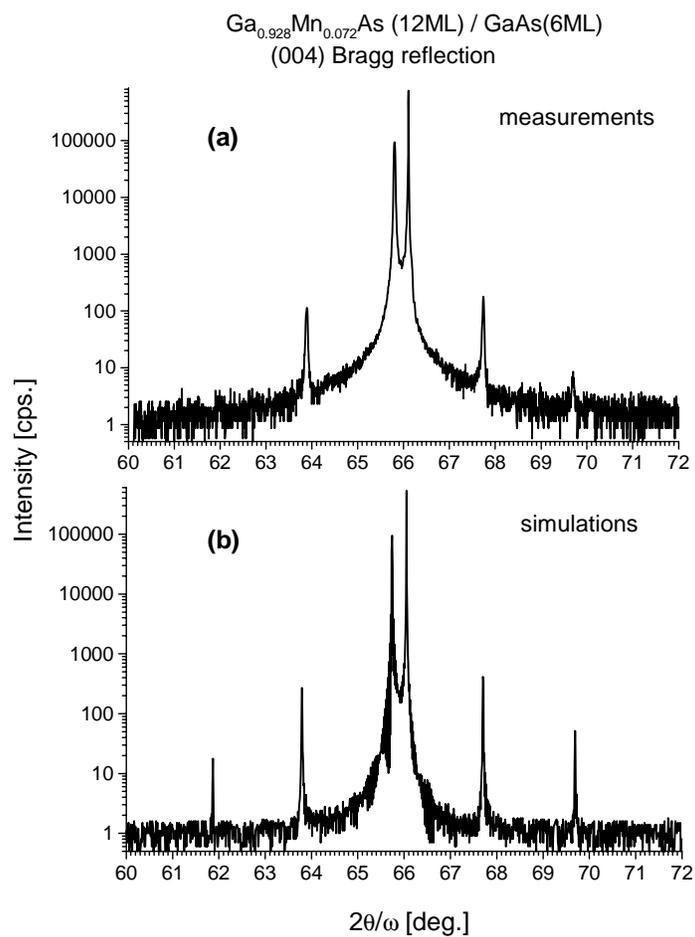

**Fig. 4.** (004) X-ray Bragg reflection for a $Ga_{0.928}Mn_{0.072}As$(12 ML) / GaAs(6 ML) x 100 superlattice: (a) – measured spectrum, (b) - calculated spectrum.

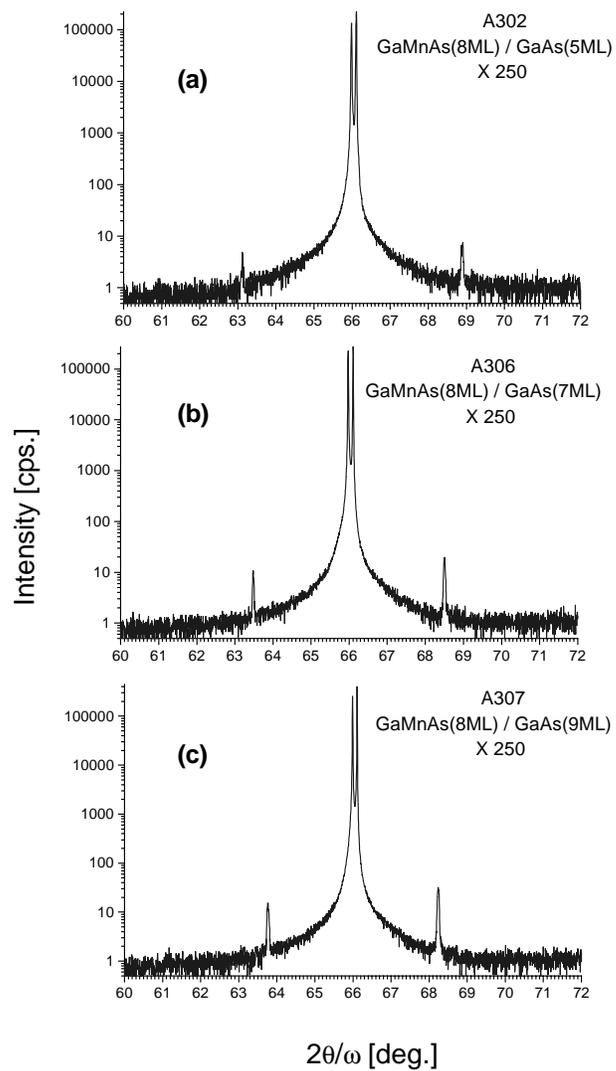

**Fig. 5.** (004) X-ray Bragg reflections for $Ga_{0.96}Mn_{0.04}As$ (8 ML) / GaAs (n ML):

(a) n = 5, (b) n = 7, (c) n = 9.

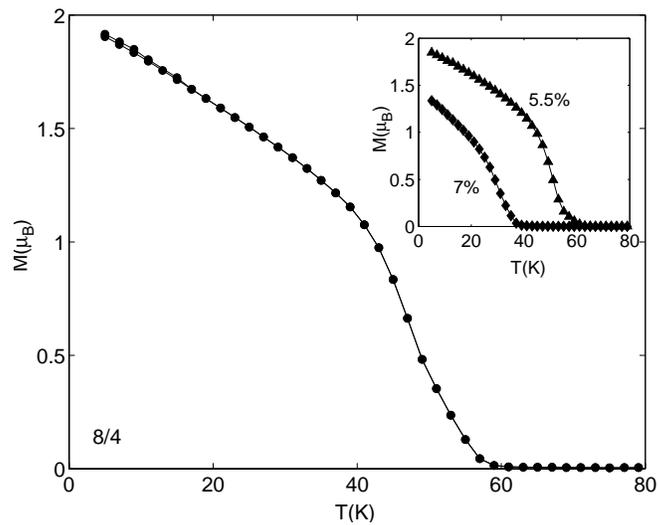

**Fig. 6.** Temperature dependence of the magnetization for a $Ga_{0.93}Mn_{0.07}As$ (8 ML) / GaAs(4 ML) x 100 SL. The inset shows results for two 0.8 _m thick reference GaMnAs samples, containing 5.5% and 7% of Mn, respectively.

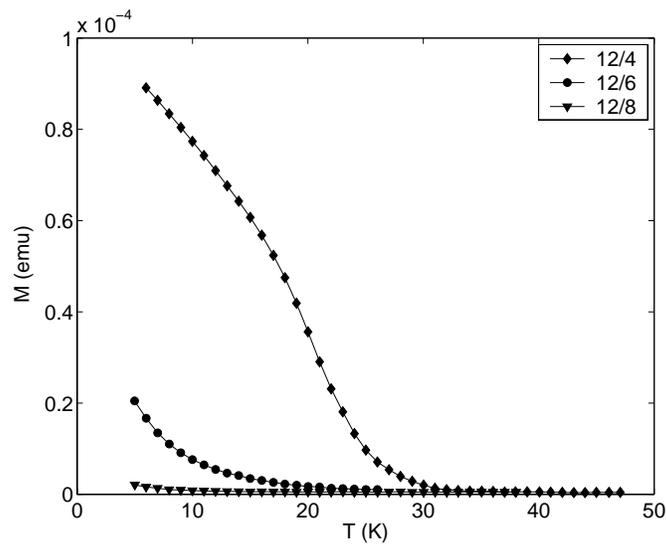

**Fig. 7.** Temperature dependence of the magnetization for $Ga_{0.96}Mn_{0.04}As$(12 ML) / GaAs(n ML) superlattices. In superlattice structures with the thickness of GaAs being larger than 10 ML, no ferromagnetic phase transition was detected.

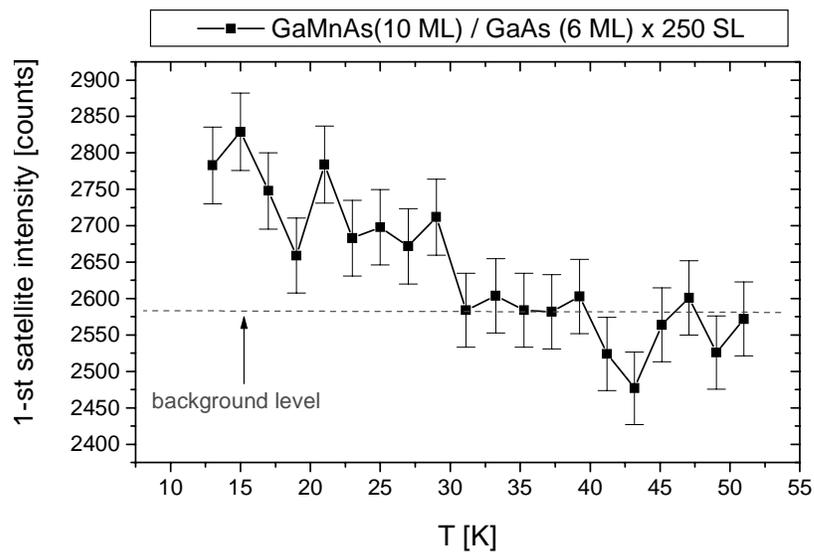

**Fig. 8.** Intensity of the 1-st satellite peak in a $Ga_{0.93}Mn_{0.07}As$(10 ML) / GaAs(6 ML) SL vs. temperature, detected in neutron diffraction experiments. An increase of the satellite intensity below $T_c$ indicates that the adjacent GaMnAs superlattice layers are ferromagnetically coupled.